# Darwinian spreading and quality thinning in even-aged boreal forest stands

Petri P. Kärenlampi[*]

Lehtoi Research, Finland
petri.karenlampi@professori.fi

[*] Author to whom correspondence should be addressed.

## Abstract

Darwinian spreading of vigor, in addition to quality distribution, is introduced in a tree growth model. The size of any tree, within an even-aged stand, is taken as a measure of an inherited productive capacity, and then combined with quality thinning. For sparse cultivation density, the result is forestry without commercial thinnings; large trees cannot be removed since they are the most productive, neither small trees because the unit price of harvesting would be large. For large cultivation density, thinning from above becomes combined with quality thinning of large trees. Darwinian spreading of growth rate may correlate with quality. Quality thinning may enhance growth rate. Such effects combined, large trees still become removed but quality thinning becomes implemented in almost all diameter classes. At financial maturity, Darwinian spreading treated with quality thinning, the weighted mean breast height diameter is in the vicinity of 20 cm within stands of high planting density, and somewhat higher with low planting density. Quality correlating with Darwinian spreading of growth, and growth rate correlating with observed quality, trees grow bigger, the maturity size becoming closer to 25 cm - bigger than in earlier studies without tree-size - dependency of vigor.

## Introduction

Properties of individuals vary, and some part of the variation is inherited [Darwin 1859, Campos et al. 2025]. Plants competing of radiation and space, some individuals dominate over others – partly due to coincidence, but partially due to their inherited capabilities.

It has been recognized that single "average" values of input parameters do not necessarily suffice as input for growth and yield models [Green and Strawderman 1996, MacFarlane et al. 2000]. However, connection of quality distributions of trees to operative stand management appears to be missing in the literature. This paper attempts to establish such a connection.

A variety of forest growth models have been established, describing the effect of environmental circumstances, as well as measurable tree characteristics to the size increment rate of trees [Monserud and Sterba 1996; Palahi et al. 2003, Trasobares et al. 2004a. 2004b, Bollandsås et al. 2008, Pukkala et al. 2009, 2010, Rosa et al. 2018]. However, surprisingly, such tree growth models discuss the expected value of the size increment of any tree, omitting any variation between trees of similar external characteristics. It is worth asking whether negligence of inherent variability in productive capacity possibly induces bias in silvicultural procedures designed.

Not only does productive capacity vary between individuals, but there also is inherent variability in qualitative characteristics. Again, most known models focus on expected values of quantities, and neglect quality variation. However, some very recent investigations have discussed quality variations between trees, as well as prospectives of quality thinning [Kärenlampi 2025a, 2025b]. In these recent studies, the quality distribution was described similarly for all trees, regardless of size and position within the stand. Quality does depend on the position of any tree within the stand. The development of suppressed trees is restricted by suppressing trees [Bollandsås et al. 2008, Groot and Hökkä 2000, Hökkä and Mäkelä 2014, Hynynen et al. 2019, Lehtonen et al. 2023]. The suppressing trees may be either older or inherently faster growing than the suppressed trees, or both.

This paper intends to include the inherent variability in productive capacity of trees into a tree growth model and clarify the corresponding effect on suitable silvicultural practices. The inherent variability can be considered in management applications only if it is measurable. This paper discusses even-aged forest stands and adopts the tree size as a measure of the individual productive capacity. The inherent variability of vigor is then combined with quality variation of trees, resulting in eventual quality thinnings.

The productive capacity and the tree quality traits may or may not correlate. It may be natural to consider growth vigor as one of the features included in the concept of quality. It also is possible that individuals with greatest vigor result in products of the highest value. As the last treatment of this study, such effects become combined, and the resulting silvicultural procedures are reported.

**Methods and Materials**

The methods applied in this study are the same as in a recent investigation [Kärenlampi 2025a] with three exceptions. The first exception is the data, as described below in detail. The other exception is the adoption of tree size as a measure of vigor, as also detailed below. Thirdly, some non-Markovian growth modelling is included.

*Applied growth model*

An inventory-based growth model [Bollandsås et al. 2008] is applied. For any tree species and diameter class, the expected diameter change rate is computed for 30-month periods. In this study, there is one deviation from the original growth model; recruitment of new trees into the smallest diameter class is omitted. The reason is that the present study discusses even-aged stands, and recruitment of new trees would render any stand non-even-aged. The physical correspondence of the even-aged stand, without natural recruitment, is a stand on soil that does not readily host natural tree seeds. Abandoned agricultural land often is one example.

*Applied financial model*

We apply a procedure first mentioned in the literature in 1967, but applied only recently [Speidel 1967, 1972, Kärenlampi 2019a, 2020a, 2020b, 2021a, 2021b, 2022a]. Instead of discounting revenues, the return rate on capital achieved as relative value increment at different stages of forest stand development is weighed by current capitalization, and integrated. Such a procedure has been shown to avoid bias resulting from approximations based on compounding equations [Kärenlampi 2025c].

The return rate on capital is the relative time change rate of value. We choose to write

$$r(t) = \frac{d\kappa}{K(t)dt} \qquad (1)$$

where $\kappa$ in the numerator considers value growth, operative expenses, and amortizations, but neglects investments and withdrawals. In other words, it is the change of capitalization on an operating profit basis. $K$ in the denominator gives capitalization on a balance sheet basis, being directly affected by any investment or withdrawal.

A momentary rate of the rate of return on capital of Eq. (1) naturally is not representative for an entire forestry system – an expected value of the return rate is needed for management considerations. As time proceeds linearly, the expected value of the return rate on capital within the rotation can be written as [Kärenlampi 2022b, 2022a]

$$\langle r(t) \rangle = \frac{\int_b^{b+\tau} \frac{d\kappa}{dt}(a)da}{\int_b^{b+\tau} K(a)da} = \frac{\int_b^{b+\tau} r(a)K(a)da}{\int_b^{b+\tau} K(a)da} \qquad (2),$$

where $a$ is stand age, $b$ is arbitrary starting age of the integration, and $\tau$ is rotation age. Along with a periodic boundary condition, Eq. (2) does not depend on the starting point of the integration.

*Variation of productive capacity between trees*

In an even-aged forest, the size of any tree obviously is a measure of the productive capacity of any tree. The variation in tree size, however, depends on other factors, along with the inherited production capacity, including spacing of trees, aspect, slope, among others. Then, the variation in tree size only partially reflects inherited production capacity. The simplest way to approximate an inherited production capacity coefficient apparently is

$$m_i = \alpha \frac{d_i}{\langle d \rangle} + (1-\alpha) \qquad (3),$$

where $m_i$ is the inherited production capacity coefficient of a tree with diameter $d_i$, $\langle d \rangle$ is the average diameter of all trees on the site, and $\alpha$ is an inheritance contribution coefficient, given the value of 0.5 throughout this study.

Introduction of the simple coefficient from Eq. (3) into the tree diameter increment rate gained from the original version of the growth model [Bollandsås et al. 2008] might have an immediate effect on the total growth rate of trees on the stand. To avoid such eventual bias, a corrected coefficient is defined as

$$n_i = m_i \frac{\Delta BA(0)}{\Delta BA(m)} \qquad (4),$$

where $\Delta BA(m)$ is stand basal area change rate with the correction coefficients $m$, and $\Delta BA(0)$ is the basal area change rate in the absence of any correction. The corrected coefficient from Eq. (4) is then used to multiply any tree diameter increment rate gained from the original version of the growth model [Bollandsås et al. 2008].

The outcome of Eq. (4) depends on the phase of stand growth where it is applied. In this study, the Equation is always applied 90 months after the initial observation of saplings in the two lowest tree diameter classes. During further stand development, the correction coefficients given by Eq. (4) evolve according to Eq. (5):

$$n_{d,t+1} = \frac{NT_{d-1,t}n_{d-1,t} + NR_{d,t}n_{d,t}}{NT_{d-1,t} + NR_{d,t}} \qquad (5),$$

where $NT_{i,j}$ is the number of trees transferring from size class $i$ to size class $i+1$ at time step $j$, and $NR_{k,l}$ is the number of trees remaining in size class $k$ at time step $l$.

As stated in the Introduction, a third procedure in the examination of the effects of Darwinian spreading of vigor is the eventual inclusion of quality in the concept of vigor. In the case of the inclusion, the correction coefficient $m$ also scales the quality parameter to be discussed below.

*Quality distributions and quality thinning*

Instead of having a constant value per trunk or per cubic meter within a class of trees of specified tree species and size, the properties vary. The expected monetary value of the assortments from any tree is determined based on the expected yield of pulpwood and sawlogs, along with the prices of the assortments [Kärenlampi 2022d]. Here, the distribution of quality is modeled as a uniform distribution between a minimum and a maximum, these being placed symmetrically with respect to the mean. Correspondingly, quality thinning results in a value correction for the remaining trees as
$j = 1+b(1-p)$ \qquad (6),
where $+-b$ are the relative limits of the uniform distribution, set as $+-0.5$ throughout this paper, and $p$ is the survival rate of trees in the quality thinning. Eq. (6), if applied, quality thinning is here applied to the first commercial thinning.

Quality thinning, however, cannot be applied on strip roads, where all the trees have to be removed. Then, after opening the strip roads, the survival rate of trees in the quality thinning is
$p=s/a$ \qquad (7),
where $s$ is the total survival rate in the harvesting, and $a$ is the survival rate after opening strip roads, but before any eventual quality thinning.

A somewhat nontrivial question is, how the value correction due to quality should be applied. Pulpwood-sized trees contain pulpwood only, regardless of their quality. However, good-quality pulpwood stems may later contain a larger proportion of sawlogs, or possibly high-value specialty logs. Let us write the expected value of the roadside price of a single trunk of specified size and species as
$$\langle P_{tree} \rangle = \langle P_{pulp} \rangle + \langle P_{saw} \rangle = \langle v_{pulp} \rangle \langle p_{pulp} \rangle + \langle v_{saw} \rangle \langle p_{saw} \rangle \qquad (8),$$

where the right-hand side quantities correspond to the expected volumes and unit prices. However, in the case of quality variation, most, if not all, of the quantities on the right-hand side vary. Introducing a statistical distribution into all these quantities would be a complicated task. An even more serious problem is that in the case of pulpwood size trees, the sawlog content is always zero, even if the tree quality obviously affects the sawlog value to be gained later.

Apparently, such problems can be avoided by assigning the quality variation into the sawlog unit price. Then, the roadside price of a trunk of specified size and species becomes

$$P_{tree} = \left\langle P_{pulp} \right\rangle + P_{saw} = \left\langle v_{pulp} \right\rangle \left\langle p_{pulp} \right\rangle + \left\langle v_{saw} \right\rangle p_{saw} = \left\langle v_{pulp} \right\rangle \left\langle p_{pulp} \right\rangle + \left\langle v_{saw} \right\rangle j \left\langle p_{saw} \right\rangle \qquad (9).$$

It is found from Eq. (9) that the quality only contributes to the roadside value if both the expected value of sawlog content and the expected value of sawlog price differ from zero, and then is proportional to both these quantities.

During further stand development after quality thinning, the correction coefficients given by Eq. (6) again evolve according to Eq. (5).

Again, as stated in the Introduction, a third procedure in the examination of the effects of quality thinning is the eventual inclusion of the tree vigor in the concept of quality. In the case of the inclusion, the correction coefficient $j$ also scales the tree diameter increment rate in the growth model.

All the modeling above is Markovian. In other words, the status change of any tree within a 30-month development step depends on the state of the tree and that of the stand in the beginning of the 30-month development step. In reality, the status change depends not only on the present status, but also on history; there often is some kind of a delay in organisms reacting to environmental changes. Here, any history-dependency of tree growth is described by averaging any diameter increment rate predicted by the growth model [Bollandsås et al. 2008] with the increment rate predicted for the preceding 30-month period.

*Example stands*

As mentioned above, the dataset used here significantly differs from the one used in the recent investigation [Kärenlampi 2025a]. Instead of using observations from never-thinned wooded stands, the growth model is here applied as early as it is applicable. This reflects greater reliance on the growth model, based on large experimental datasets, in relation to approximations of historical development resulting in presently measurable observables. The growth model is applied to young stands where breast-height diameters are 6–11 cm, which was achieved at 15–20 years of age, depending on tree species [Vuokila 1956,1960, Vuokila and Väliaho 1980, Raulo 1977, Oikarinen 1983]. Such an approach, however, requires the investigation of a variety of alternative stem counts cultivated and retained in eventual young stand tending. We choose to investigate stem counts 1200/ha and 2400/ha.

As the growth model is not applicable to young seedlings, the stand development until breast-height diameters 6–11 cm must be approximated some other way. For this period, an exponential volume growth, as well as an exponential value increment is assumed. Further on, volumetric growth is produced using the growth model [Bollandsås et al. 2008, Kärenlampi 2019b,c], and value growth also using the assortment yield model [Kärenlampi 2020d, 2021a, 2022d, Mehtätalo 2002, Laasasenaho 1982].

## Results

The expected value of the return rate on capital, as a function of rotation age, is shown in Figs. 1a to 1c. Figure 1a shows a reference case, where both Darwinian spreading of vigor and quality variation are neglected. Figure 1b shows a case where Darwinian spreading of vigor contributes to diameter growth rate, and quality thinning contributes to the quality parameter according to Eq. (6). Figure 1c shows a case where Darwinian spreading of vigor contributes to diameter growth rate and the quality parameter, and quality thinning contributes to both quality and growth rate.

In Fig. 1a, the rotations are longer than in Fig. 1b, but the achievable rates of return do not differ much. The reason for the longer rotations in Fig. 1a is that stands are thinned only from above, whereas in Fig. 1b there is partial quality thinning. An exception are the sparsely cultivated pine and birch stands – the rotation lengths are the same, as such stands do not experience any thinning in either Figure. Sparsely cultivated spruce stand is thinned from above in Fig. 1a, and the rotation is correspondingly extended, whereas there is no thinning of the similar stand in Fig. 1b.

It is of interest, why the return rates are not higher in Fig. 1b than in Fig. 1a, even if variation in vigor and quality are considered in Fig. 1b. The reason is in the negligence itself: thinning from above retains small trees of low vigor, but the vigor variation is neglected, resulting in an overestimation of the growth rate of the small trees in Fig. 1a.

In Figure 1c, in stark contrast to Figs. 1a and 1b, all stands experience quality thinning. The return rates on capital are greater than in Figs. 1a and 1b. In the case of sparsely cultivated stands not thinned in Figs. 1a and 1b, such stands show longer rotations in Fig. 1c, as the thinnings are still implemented predominantly from above.

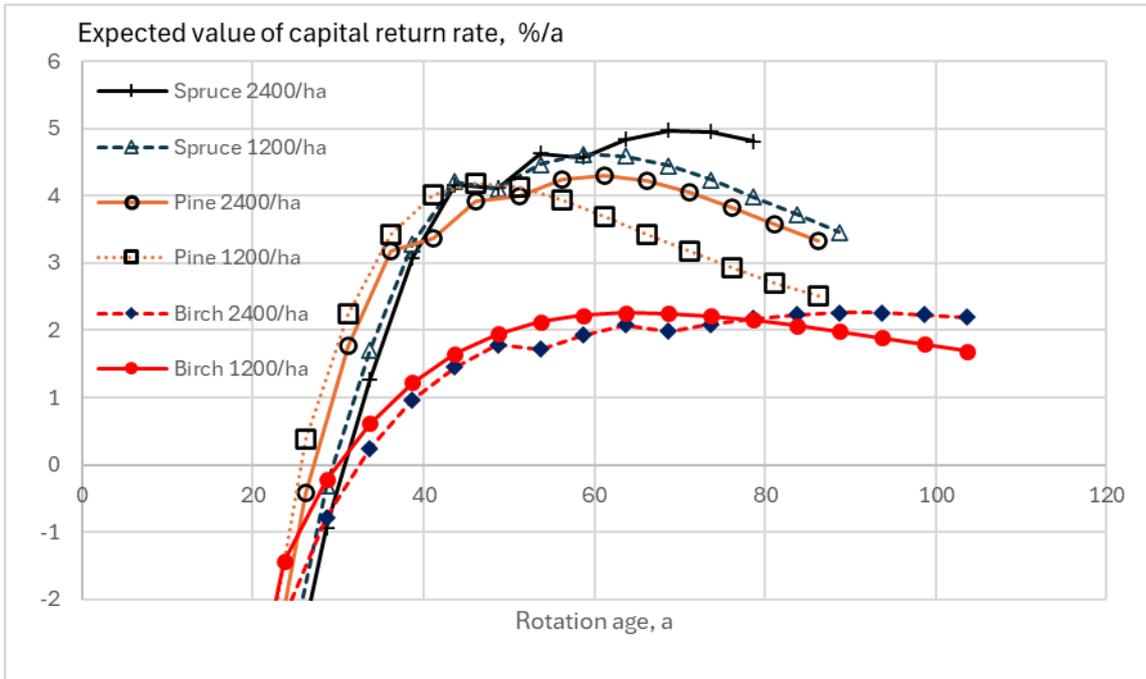

Fig. 1a. Expected value of return rate on capital as a function of rotation age, both Darwinian spreading and quality variation neglected.

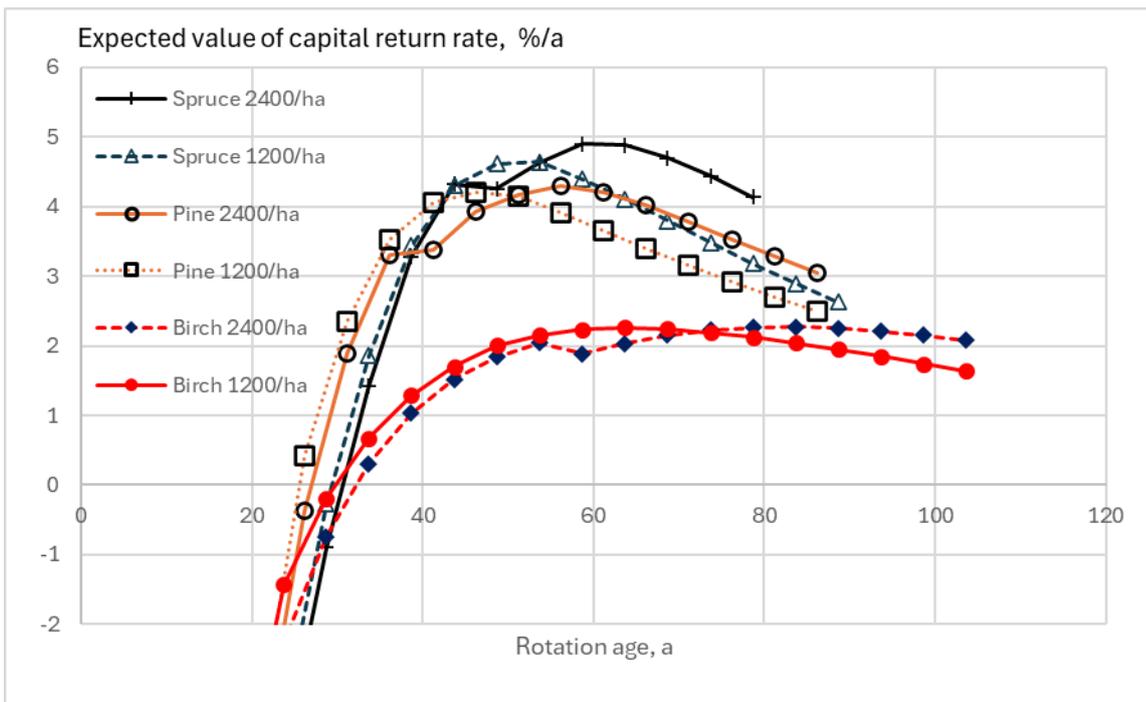

Fig. 1b. Expected value of return rate on capital as a function of rotation age, quality thinning applied on Darwinian spreading.

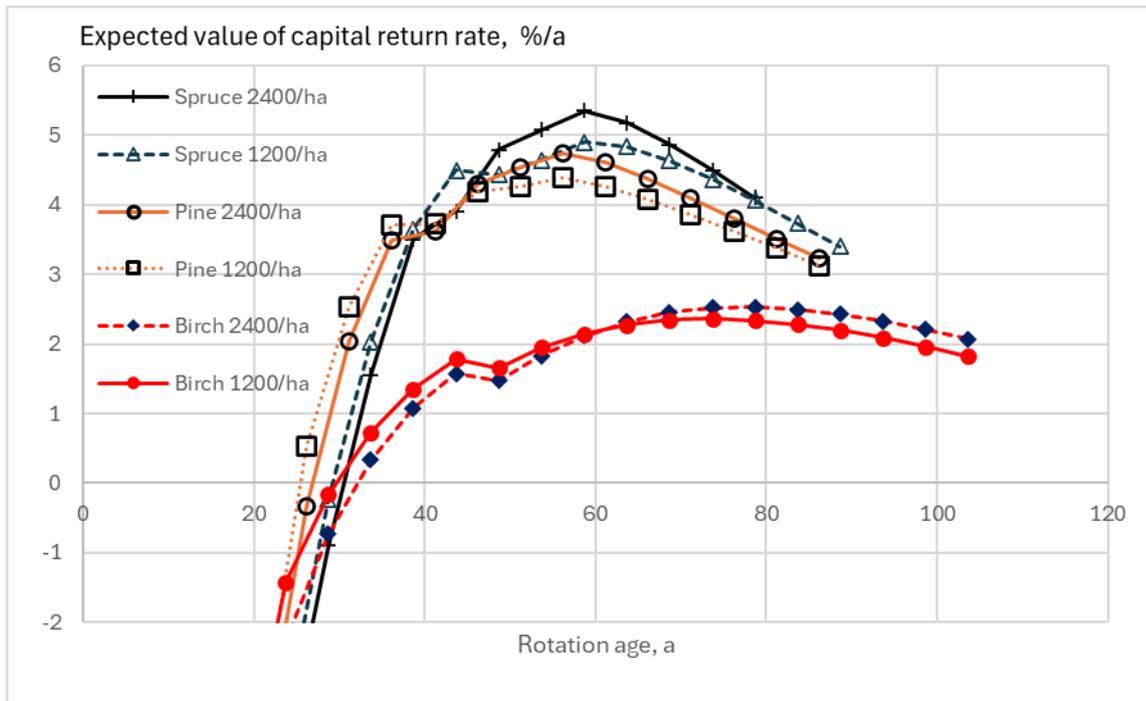

Fig. 1c. Expected value of return rate on capital as a function of rotation age, quality correlating with Darwinian spreading of growth, growth rate correlating with observed quality.

The basal area-weighed mean breast-height diameter as a function of stand age is shown in Figs. 2a to 2c. The drawings are terminated at financial maturity. Figure 2a again shows the reference case, where both Darwinian spreading of vigor and quality variation are neglected, and Figure 2b shows a case where Darwinian spreading of vigor contributes to diameter growth rate, and quality thinning contributes to the quality parameter according to Eq. (6). Figure 2c shows a case where Darwinian spreading of vigor contributes to diameter growth rate and the quality parameter, and quality thinning contributes to both quality and growth rate.

In Fig. 2a, the rotations are again longer than in Fig. 2b, but the average tree diameters at financial maturity do not differ much. Most of the stands show an average maturity diameter in the vicinity of 20 cm. The sparsely cultivated spruce stands shown greater maturity diameters, and larger in the case of no thinning from above in Fig. 2b.

In Figure 2c, rotations are longer and trees grow bigger, in a few cases reaching the vicinity of 25 cm.

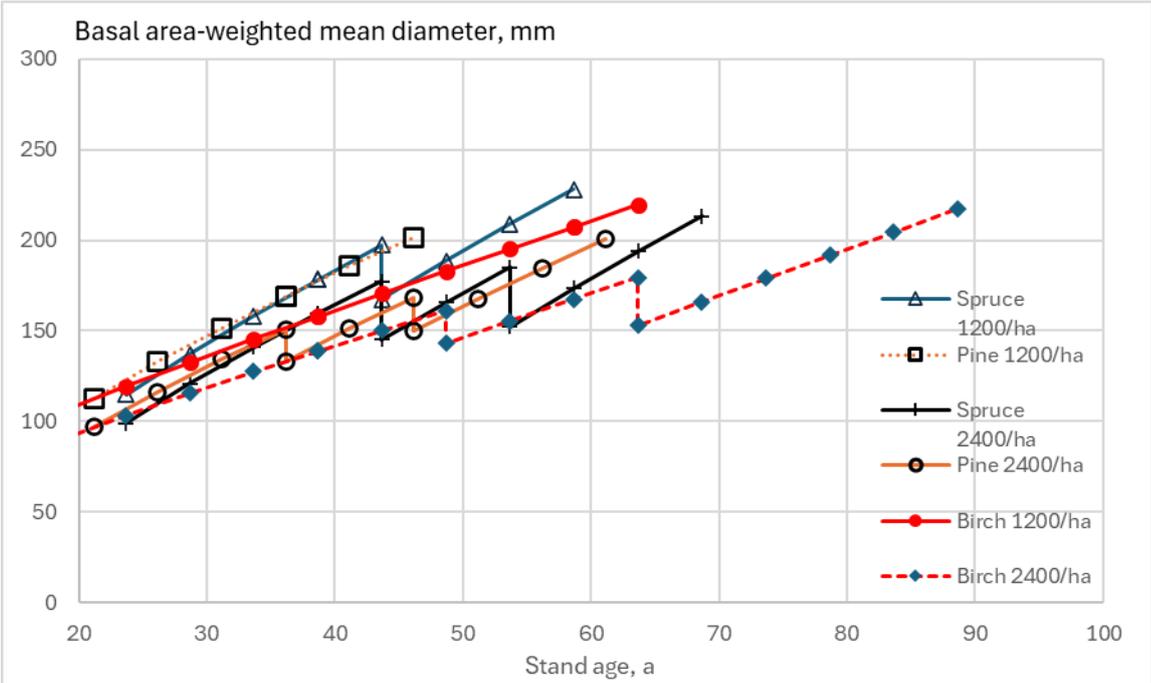

Fig. 2a. Basal area-weighted average diameter as a function of stand age, both Darwinian spreading and quality variation neglected.

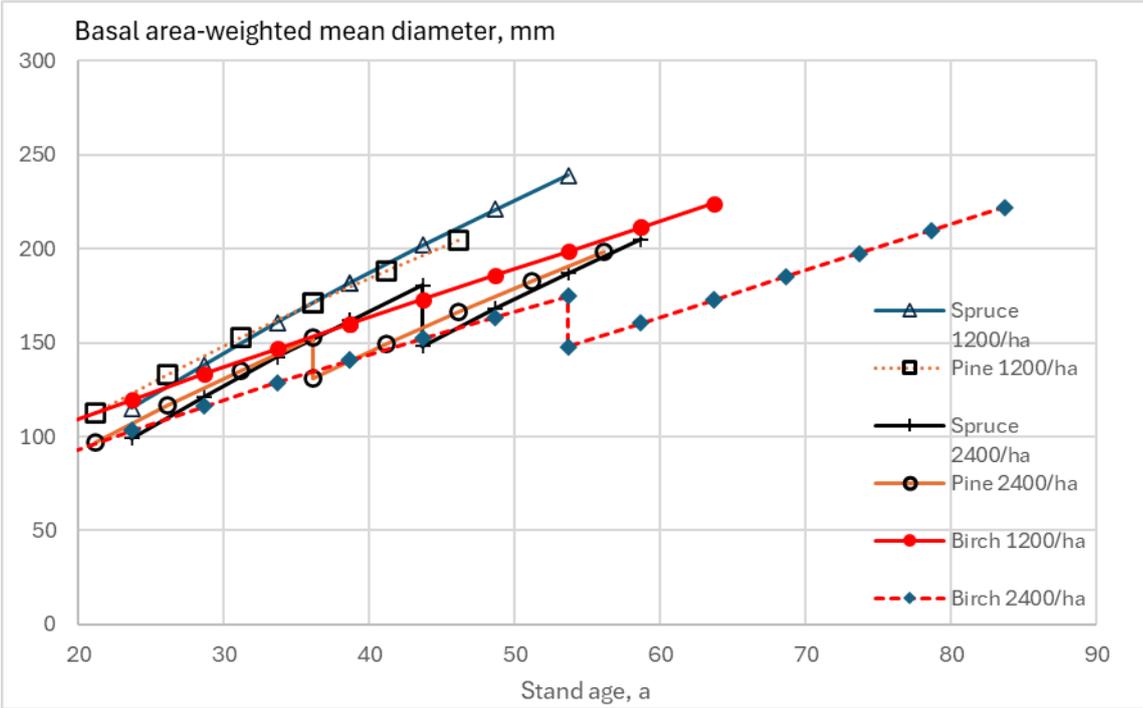

Fig. 2b. Basal area-weighted average diameter as a function of stand age, quality thinning applied om Darwinian spreading.

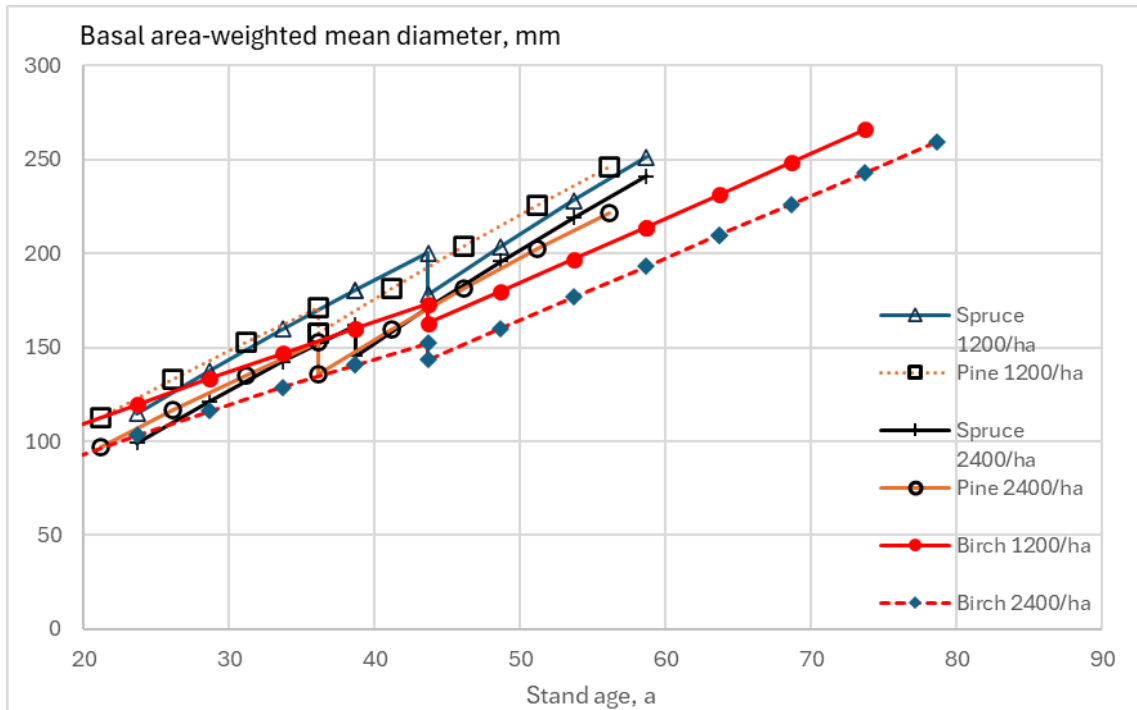

Fig. 2c. Basal area-weighted average diameter as a function of stand age, quality correlating with Darwinian spreading of growth, growth rate correlating with observed quality.

The proportion of basal area removed in thinning, by diameter class, is shown in Figs. 3a to 3c. Figure 3a again shows the reference case, where both Darwinian spreading of vigor and quality variation are neglected, and Figure 3b shows a case where Darwinian spreading of vigor contributes to diameter growth rate, and quality thinning contributes to the quality parameter according to Eq. (6). Figure 3c shows a case where Darwinian spreading of vigor contributes to diameter growth rate and the quality parameter, and quality thinning contributes to both quality and growth rate.

In Fig. 3a, thinnings are implemented from above, and small trees are removed only from striproads. Pine and birch stands with sparce cultivation density remain unthinned. In Fig, 3b, thinnings are still implemented from above, but quality thinning extends to a few diameter classes – however not to the smallest. Sparsely cultivated stands of all three tree species remain unthinned.

In Figure 3c, thinnings are still implemented from above, but quality thinning now extends to all diameter classes, with one exception.

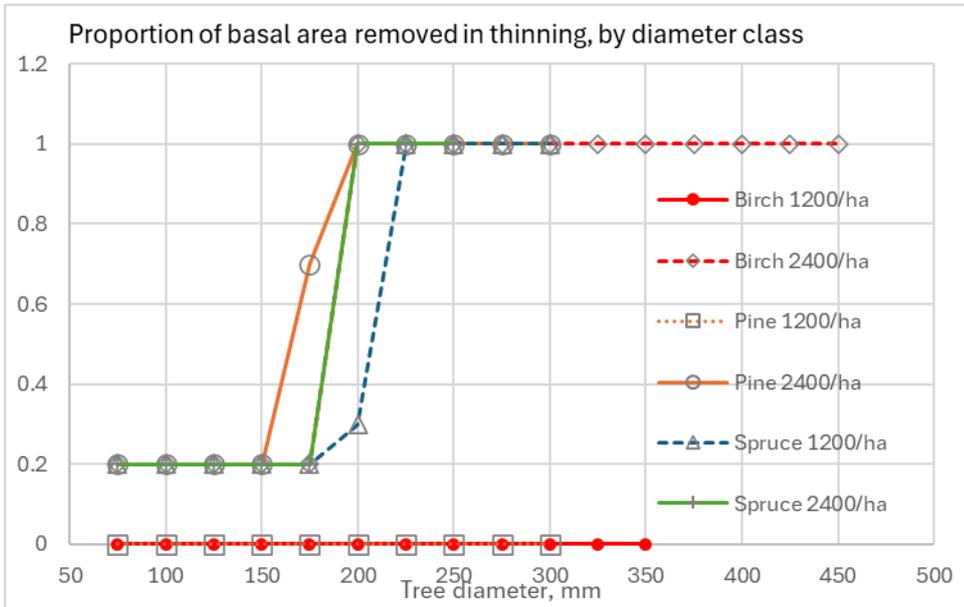

Fig. 3a. Proportion of basal area removed in thinning, both Darwinian spreading and quality variation neglected.

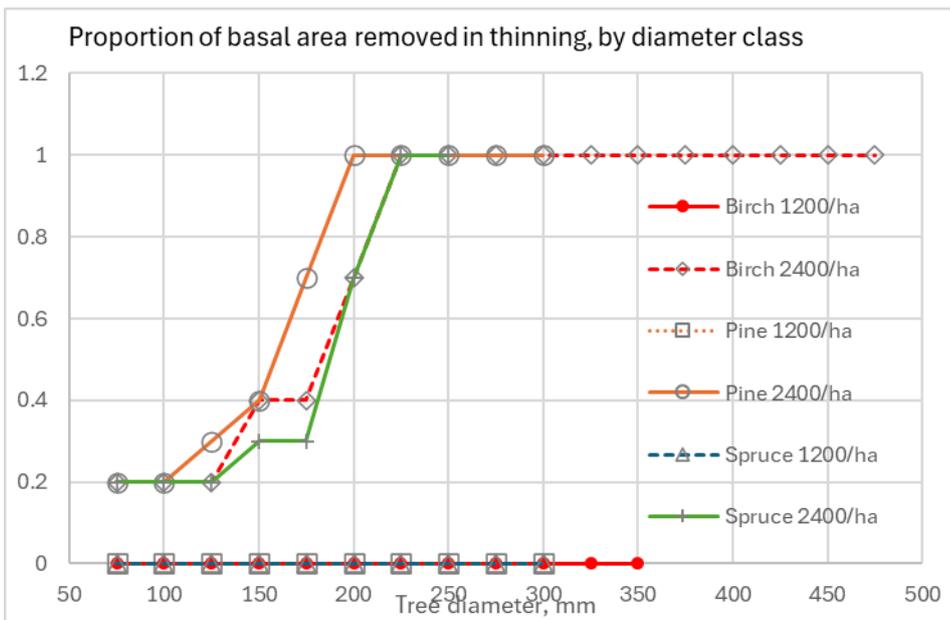

Fig. 3b. Proportion of basal area removed in quality thinning, applied on Darwinian spreading, per 250 mm diameter class.

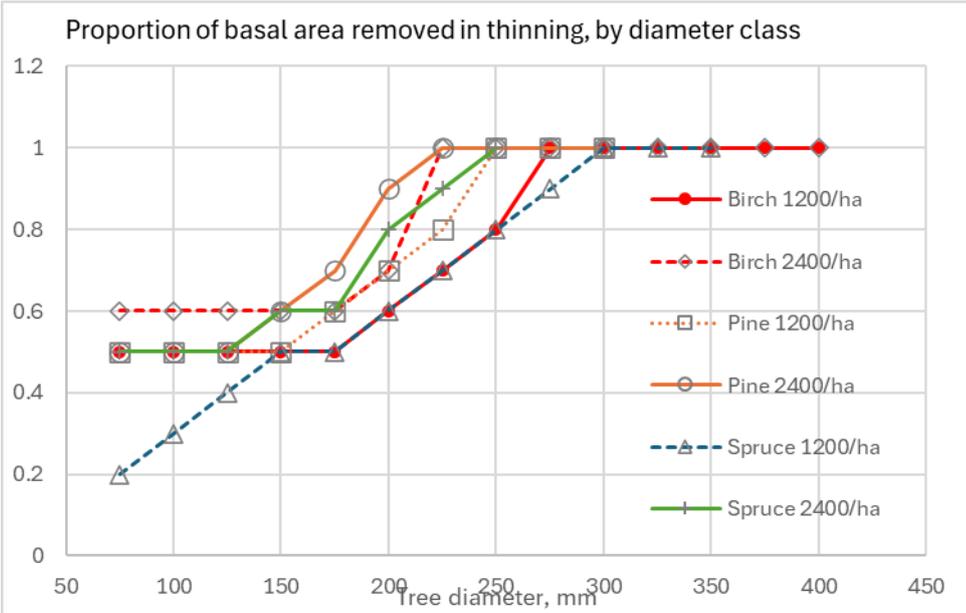

Fig. 3c. Proportion of basal area removed in quality thinning, per 250 mm diameter class, quality correlating with Darwinian spreading of growth, growth rate correlating with observed quality.

The basal area and stem count triggering thinning, as well as those retained after thinning, are shown in Figs. 4a to 4c. Figure 4a again shows a reference case, where both Darwinian spreading of vigor and quality variation are neglected, and Figure 4b shows a case where Darwinian spreading of vigor contributes to diameter growth rate, and quality thinning contributes to the quality parameter according to Eq. (6). Figure 4c shows a case where Darwinian spreading of vigor contributes to diameter growth rate and the quality parameter, and quality thinning contributes to both quality and growth rate.

In Fig. 4a, thinnings from above remove about half of the basal area, and less than half of the stemcount. Pine and birch stands with sparce cultivation density remain unthinned. In Fig, 4b, the triggering and the severity of the thinnings differs only slightly. Sparsely cultivated stands of all three tree species remain unthinned. In Figure 4c, all stands are severely thinned, as most of the basal area and at least half of any stem count is removed.

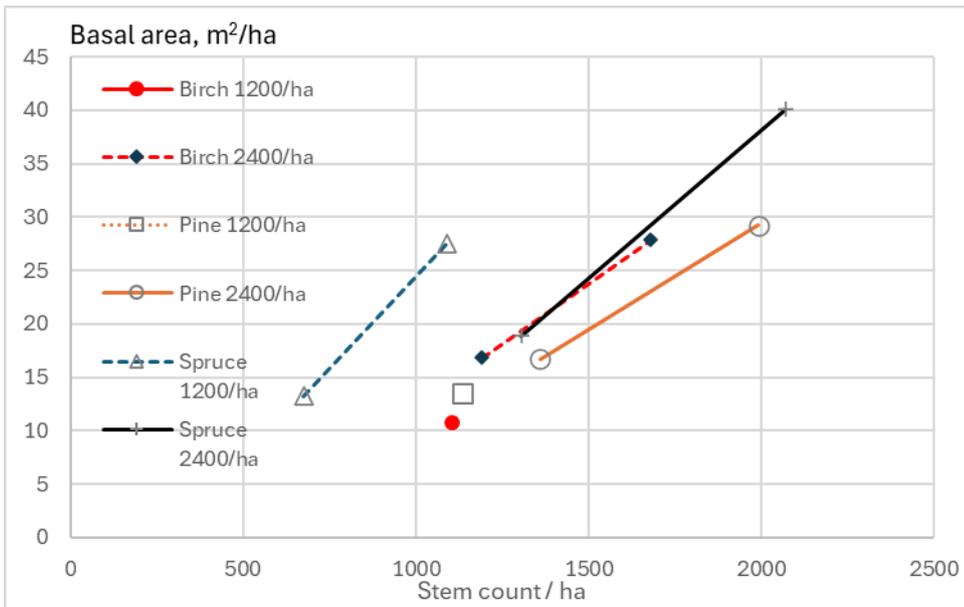

Fig. 4a. Change in basal area and stem count per hectare in thinning, both Darwinian spreading and quality variation neglected.

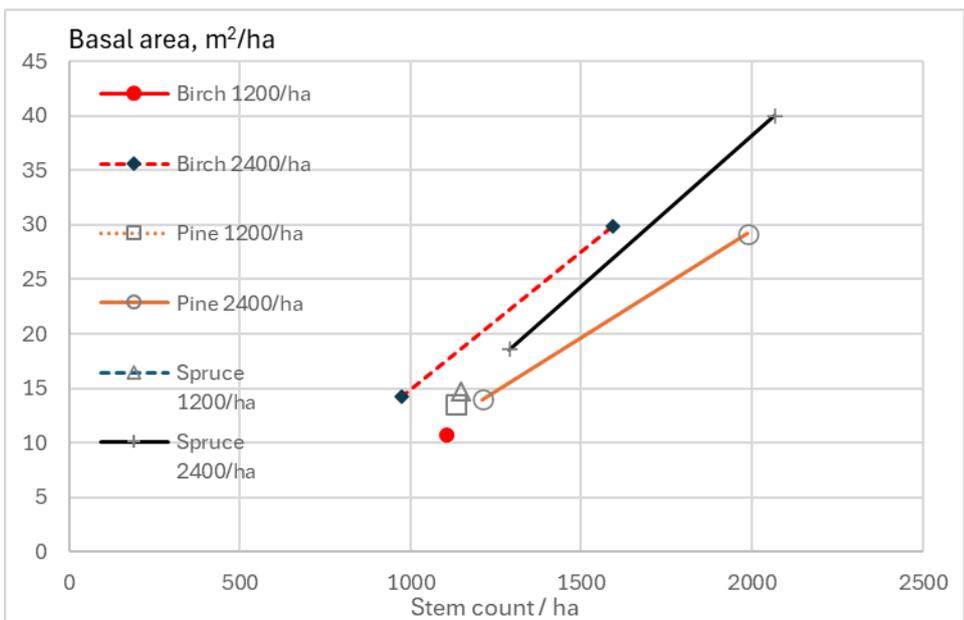

Fig. 4b. Change in basal area and stem count per hectare in quality thinning, applied on Darwinian spreading.

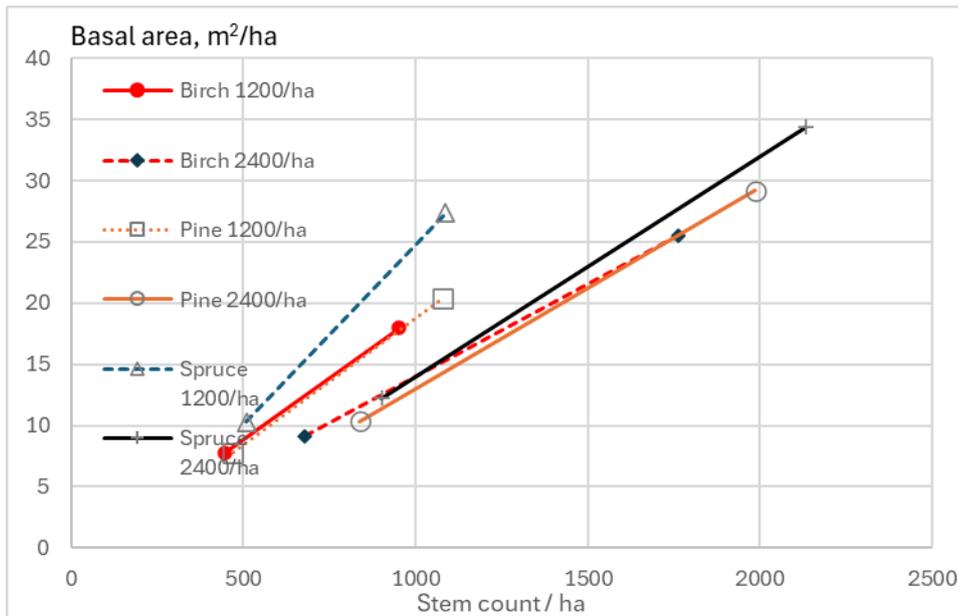

Fig. 4c. Change in basal area and stem count per hectare in quality thinning, quality correlating with Darwinian spreading of growth, growth rate correlating with observed quality.

## Discussion

A necessary question is, how the results of this paper may have been affected by the parameter values selected. Firstly, the parameter $b$ in Eq. (6) renders the roadside unit value of the worst sawlogs to that of pulpwood in Eq. (9) and thus has an economic justification. On the other hand, the parameter $\alpha$ in Eq. (3) is merely an approximation.

The effect of changing $\alpha$ in Eq. (3) was hereby investigated by changing its value to 0.7, instead of 0.5. The return rates on capital increased slightly compared to Fig. 1c, as well as the average diameters at financial maturity in comparison to Fig. 2c. However, the changes were very small, indicating that the present results are not sensitive to the value of this parameter.

Interestingly, in some cases, the reference setup, where both Darwinian spreading and quality variation were neglected (Fig. 1a), resulted in larger return rates on capital than the model including both effects (Fig. 1b). An obvious reason is in the negligence of the Darwinian spreading itself. Figs. 1a and 3a correspond to thinning from above, retaining small trees. Within an even-aged stand, the small trees have inherently lower production capacity, which has been neglected in Fig. 1a.

Another interesting observation is that in Fig. 1b, all the sparsely cultivated stands remained unthinned. Obviously, large trees cannot be removed since they are the most productive, neither small trees because the unit price of harvesting would be large.

It is of interest to compare the results of this study to those of a previous one, which discussed the effects of quality thinning, but Darwinian spreading was not included [Kärenlampi 2025b]. In other words, tree size was not used as a measure of vigor, and the stands were not strictly even-aged. Specifically, Figures 1c, 2c, 6c, and 7c of [Kärenlampi 2025b] can be compared with Figs. 1c, 2c, 3c and 4c of the present study. Somewhat higher capital return rates appear in Fig. 1c of [Kärenlampi 2025b], in comparison to 1c of this study. A possible reason for the difference is that recruitment of new saplings occurred in [Kärenlampi 2025b], unlike in this study. In half of the cases investigated, rotations are longer in [Kärenlampi 2025b].

Figure 2c, in comparison to the corresponding Figure of [Kärenlampi 2025b], shows that trees grow bigger in the present study, even if rotations are shorter. An obvious reason is that in the present study, growth vigor is scaled according to the observed tree size, and correspondingly, large trees grow faster. It is worth noting that all the results of tree size at financial maturity are smaller than present instructions applied on the relevant area [Metsänhoidon 2025].

The thinning schedules in Fig.6c of [Kärenlampi 2025b] and 3c this study do not vary much, however small diameter classes are somewhat more intensively thinned in this study. This obviously relates to removal of trees with the slowest growth rate. A comparison of Figs. 7c of [Kärenlampi 2025b] and 4c this study indicates that thinnings are triggered, as well as terminated, at higher stem counts in [Kärenlampi 2025b]. This is probably related to the recruitment of new trees in [Kärenlampi 2025b], unlike in this study.

In addition to the Markovian treatments presented above, some non-Markovian treatments were conducted in this study. Instead of depending only on the state of any stand at the initiation of any growth step, the growth reactions were made to depend on stand history. This was implemented by averaging the present growth parameter value with the one of the preceding growth period, and using the average in the computation. The non-Markovian versionresulted in greater return rates, in comparison to the Markovian version. This indicates that delayed crowding dominated over delayed recovery after thinning. In other words, the introduced history-dependence effect is asymmetric. One might ask whether this corresponds to reality.

As forest thinning is a relatively rare event, most sites sampled in the systematic inventory [Bollandsås et al. 2008] have not been recently thinned. Correspondingly, the crowding effect along with forest growth may be reasonably well described by the original Markovian model. On the other hand, there obviously is a delayed response to increased space after thinning, and such circumstances are not well represented in systematic inventory. Consequently, a non-Markovian modification of the growth model possibly should be considered mostly in the context of thinnings. On the other hand, systematic inventory has been containing some stands thinned recently. Then, the Markovian model probably slightly underestimates growth apart from thinnings, and overestimates it during post-thinning recovery.

It is worth asking, what are the mechanisms affecting the relative value increment rate and correspondingly return rate on capital [cf. Kärenlampi 2025a, Kärenlampi 2025b]. Firstly, the steepness of value time gradients contributes. Secondly, phase transitions contribute to the steepness of the value gradients. Thirdly, the effects of eventual quality thinning evolve as quality is inherited from smaller diameter classes into larger along with tree growth (Eq. (5)). Such effects can be investigated by plotting the relative value increment rate for different tree species and diameter classes, even if the integration of such measures into expected values of return rate on capital is far from straightforward. For a more detailed discussion into the mechanisms of value development, the reader is referred to the recent investigations [Kärenlampi 2025a, Kärenlampi 2025b].

Fig. 4c proposes heavy thinnings. Heavy thinnings obviously enhance the risk of snow and wind damage [Cameron 2002, Cremer et al. 1982, Persson 1972, Valinger et al. 1993]. Heavy thinnings should be avoided on exposed sites. This can be practically implemented by replacing any heavy thinning with two gentle thinnings, a few years apart, giving the remaining trees time to adapt. Such an arrangement may not be adequate on seriously exposed sites [Cameron 2002].

Fig. 1 indicates that the rate of return on capital on birch stands is much lower than that on conifer stands, and the diameter develops much more slowly along with time (Fig.2). This does not correspond well to observations of silver birch (*Betula pendula*) collected from Finland [Raulo 1977, Oikarinen 1983]. Obviously, the inventory data from Norway has mostly contained common birch (*Betula pubescens*), whereas the confer species are the same [Vuokila 1956, 1960, Vuokila and Väliaho 1980].

**Acknowledgements**

The author declares that no competing interests exist.

This work was partially funded by Niemi foundation. The funder had no role in study design, data collection and analysis, decision to publish, or preparation of the manuscript.